\documentclass[journal]{IEEEtran}
\usepackage{cite}
\usepackage{amsmath,amssymb,amsfonts}
\usepackage{graphicx}
\usepackage{textcomp}
\usepackage{xcolor}
\usepackage{epstopdf}
\usepackage{algorithm}
\usepackage{algorithmicx}
\usepackage{algpseudocode}
	
\begin{document}

\title{An Adaptive Interpolation Scheme for Wideband Frequency Sweep in Electromagnetic Simulations}

\author{Kai~Zhu,~\IEEEmembership{Graduate~Student~Member,~IEEE,}
        Jinhui~Wang,~and~Shunchuan~Yang,~\IEEEmembership{Member,~IEEE}
    
\thanks{Manuscript received xxx; revised xxx.}
\thanks{This work was supported in part by the National Natural Science Foundation of China through Grant 61801010, in part by Pre-Research Project through Grant J2019-VIII-0009-0170, and Fundamental Research Funds for the Central Universities. \textit{(Corresponding author: Shunchuan Yang)}}
\thanks{K. Zhu, and J. Wang are with the School of Electronic and Information Engineering, Beihang University, Beijing, 100083, China. (e-mail: zhukai7@buaa.edu.cn, bhexwjh@buaa.edu.cn)}
\thanks{S. Yang is with the Research Institute for Frontier Science and the School of Electronic and Information Engineering, Beihang University, Beijing, 100083, China. (e-mail: scyang@buaa.edu.cn)}
}

\markboth{Journal of \LaTeX\ Class Files}%
{Shell \MakeLowercase{\textit{et al.}}: Bare Demo of IEEEtran.cls for IEEE Journals}

\maketitle

\begin{abstract}
An adaptive interpolation scheme is proposed to accurately calculate the wideband responses in electromagnetic simulations. In the proposed scheme, the sampling points are first carefully divided into several groups based on their responses to avoid the Runge phenomenon and the error fluctuations, and then different interpolation strategies are used to calculate the responses in the whole frequency band. If the relative error does not satisfy the predefined threshold in a specific frequency band, it will be refined until the error criteria is met. The detailed error analysis is also presented to verify the accuracy of the interpolation scheme. At last, two numerical examples including the antenna radiation and the filter simulation are carried out to validate its accuracy and efficiency.
\end{abstract}


\IEEEpeerreviewmaketitle

\section{Introduction}

\IEEEPARstart{W}{ith} the fast development of high-speed electronic systems in the past decades, the demand for accurate and efficient electromagnetic simulations keeps increasing. The finite element method (FEM) \cite{FEM} and the method of moment (MOM) \cite{MOM} in the frequency domain have been widely used in electrical engineering applications, such as antenna design \cite{Antenna}, passive component design \cite{Passive2}. Unlike the finite-difference time-domain (FDTD) \cite{FDTD} methods, which can obtain a wideband response through a single simulation, the frequency sweep is required if the wideband responses are desired, such as microwave imaging \cite{Imaging}, parameter extraction in integrated circuit design \cite{Integrated}.
 
However, the wideband frequency responses usually show large and sharp variations and fluctuations, as shown in Fig. 1(a), which is the S parameter of a waveguide from 1.7 GHz to 1.9 GHz. To obtain the wideband responses, an accurate interpolation scheme is preferred since only a small number of discrete frequency points are required and significant efficiency improvement can be obtained. However, the traditional interpolation approaches usually suffer from numerical instability since many discontinuities may exist in the interested frequency band, and the increasing degree of the interpolation polynomial leads to the so-called Runge phenomenon \cite{Runge}.  It can be suppressed if the Chebyshev sampling scheme is applied \cite{Order}. However, the Chebyshev points with high non-uniformity may not be desirable in many applications. 

The model order reduction (MOR) technique can reduce the computational complexity in electromagnetic simulations. One typical approach is the asymptotic waveform evaluation (AWE) \cite{AWE} based on the Taylor series and the Pade approximation. In \cite{MOR}, the multipoint MOR was proposed to reduce the number of basis functions. However, the MOR methods suffer from accuracy issue and ill-conditioned matrix calculation \cite{AWE1}.

In this letter, an adaptive interpolation scheme is proposed to accurately and efficiently predict the wideband responses in electromagnetic simulations. A small number of frequency points in the frequency band is first selected, and their responses are calculated through field solvers. Then, we adaptively cast the sampling points into several groups to avoid numerical instability due to discontinuities. The Lagrange interpolation polynomial in each group is then constructed. If the relative error is larger than the predefined threshold in a specific frequency band, more sampling points are adaptively selected in the corresponding band until convergent results are obtained. The main merit of the proposed scheme is independent of the complexity of the matrix reduction and applicable for extremely complicated wideband responses as shown in the numerical results. 

This paper is organized as follows. In Section II, the proposed adaptive interpolation scheme is detailed presented. In Section III, its rigorous error analysis is discussed. Two numerical examples are carried out to demonstrate its effectiveness in Section IV . Finally, some conclusions are drawn in Section V.

\section{Methodology}
\subsection{Numerical Issues in the Wideband Interpolation}
To obtain accurate responses in a wide frequency band, the sampling points can be divided into several groups to obtain an effective interpolation scheme as shown in Fig. 1(a). Since p5 is an extreme point, an ideal division is that p1-p5 is casted into a group, and p5-p9 as another group. Then, the polynomial interpolation can be used to calculate the responses in the whole frequency band. However, it is not always applicable for complex scenarios. There are three issues in the practical applications: (1) If there are too many points in a group and the polynomial interpolation is used, the Runge phenomenon may happen. (2) Several extreme points are adjacent to each other when insufficient sampling occurs. (3) The extreme points may not be sampled and other points may be erroneously regarded as the extreme points. To overcome these issues, an adaptive sampling scheme is proposed in this section.

To make our explanation clear, the $i$th group denotes ${{X}_{i}}$, which has ${{N}_{i}}$ sampling points, and the distance between the $i$th and $(i-1)$th groups is ${{D}_{i}}$. ${{x}_{app}}$ is the desired frequency point and its response is ${{y}_{app}}$. The non-single group is defined as a group including more than one points, and the single group contains only one isolated or extreme point. ${{M}_{ns}}$ and ${{M}_{s}}$ are the number of non-single and single groups, respectively. The single group closest to ${{x}_{app}}$ is ${{X}_{near}}$, ${{X}_{near+1}}$ and ${{X}_{near-1}}$ denote the latter and former groups, respectively. ${{x}_{app-1}}$ is the previous point of ${{x}_{app}}$. The superscript $i$ indicates the $i$th point in the group.

\begin{figure}[H]
	\begin{minipage}[h]{0.48\linewidth}
		\centering
		\centerline{\includegraphics[width=1.80in]{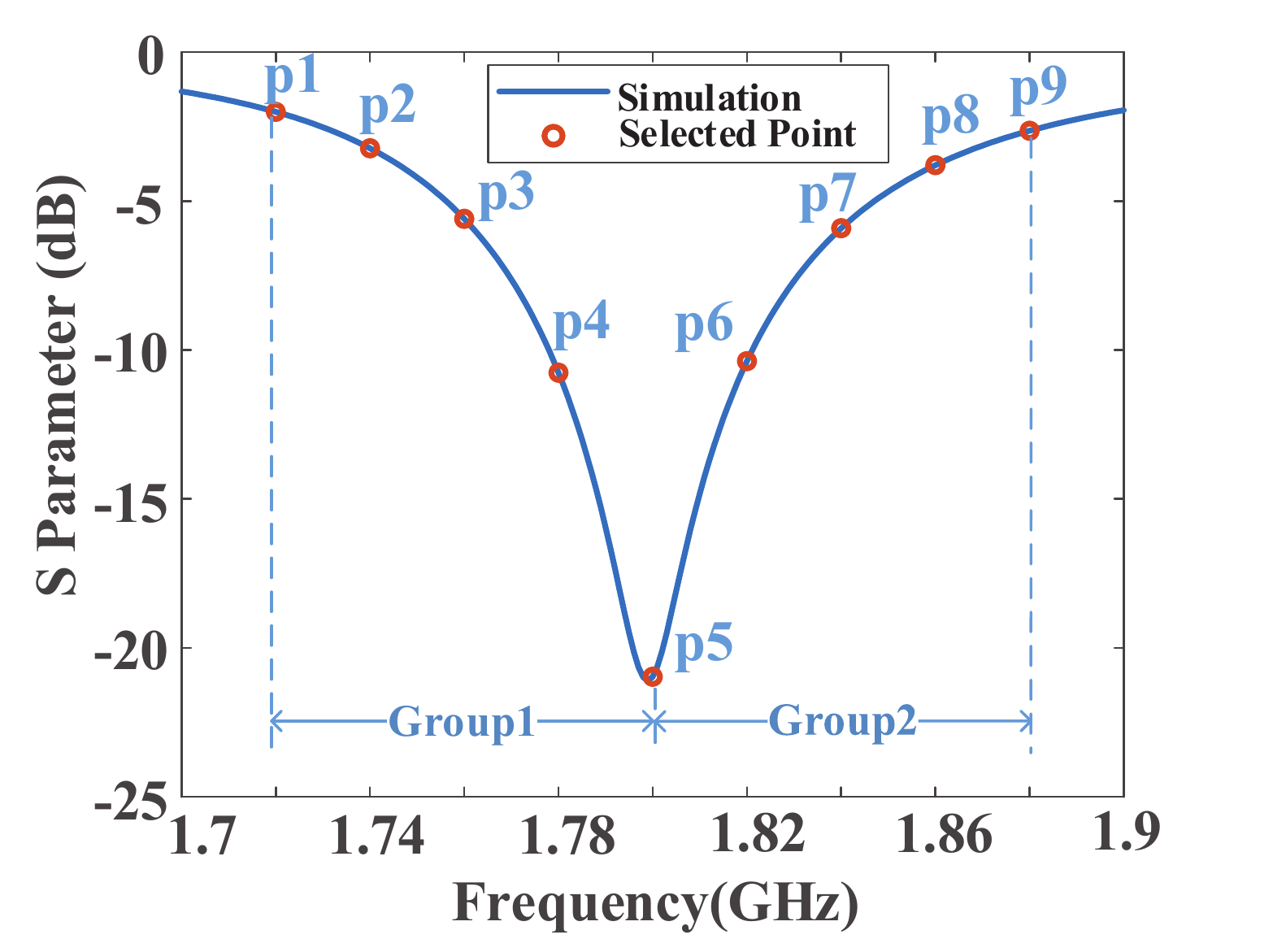}}		
		\centerline{(a)}
		\label{FIG1A}
	\end{minipage}
	\hfill
	\begin{minipage}[h]{0.48\linewidth}\label{FIG1B}
		\centerline{\includegraphics[width=1.80in]{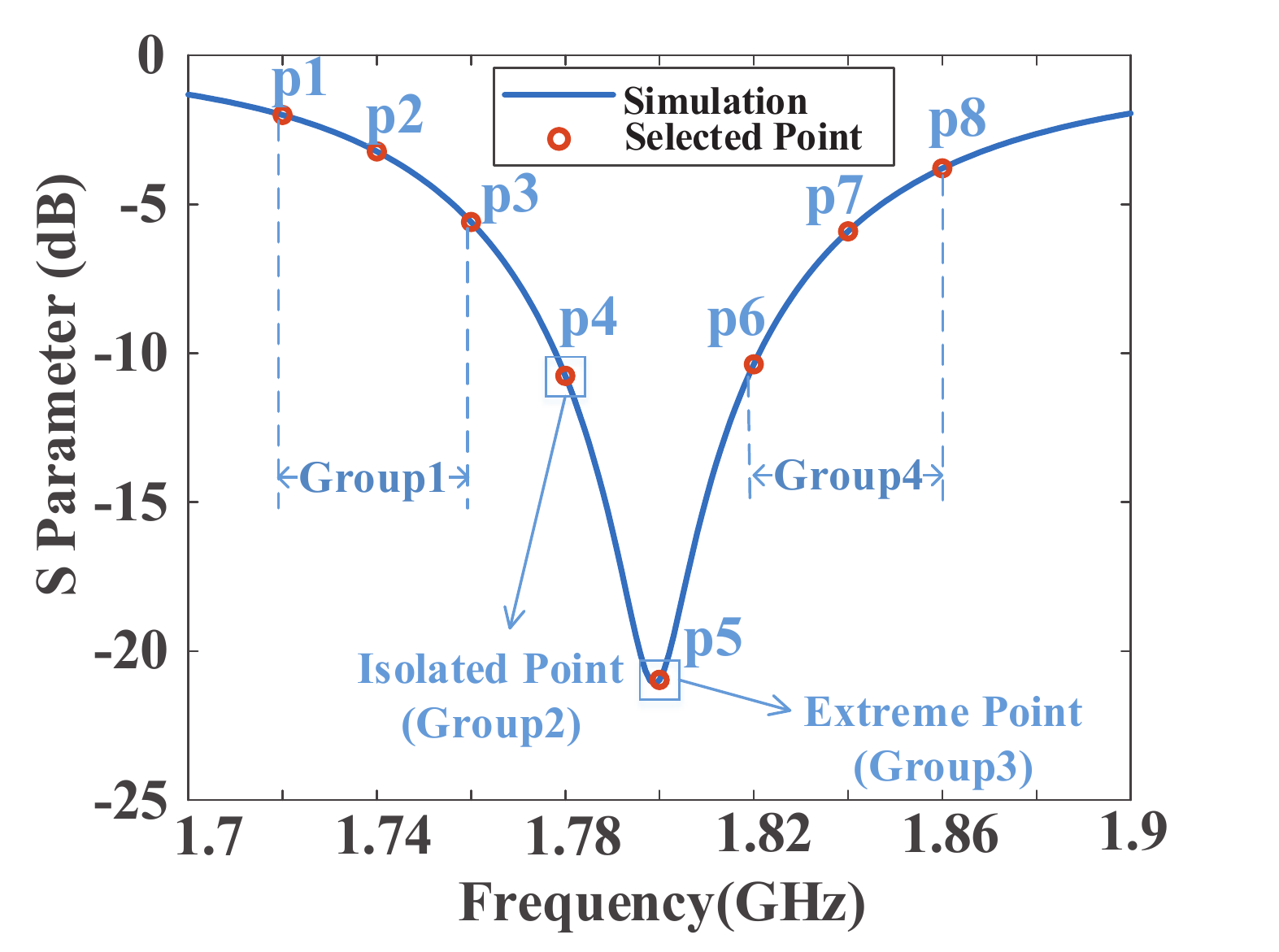}}
		\centerline{(b)}
	\end{minipage}
	\caption{Illustration of selecting the sampling points to calculate the S parameter in the frequency band for a waveguide, (a) the ideal group subdivision scheme and (b) the proposed group subdivision scheme.}
	\label{fig_6}
\end{figure}

\subsection{The Proposed Adaptive Interpolation Scheme}
The proposed interpolation scheme can be divided into four phases. First, the responses of the uniform sampling points are calculated at the beginning, which are then divided into several groups according to the extreme points. Second, the Lagrange interpolation is applied in each group, and the frequency response in the whole frequency range are analyzed. Third, the frequency band is divided into several parts, the relative error in each part is evaluated. Finally, more sampling points are carefully selected in the specific frequency bands, where the relative errors are larger than the predefined values, and then they are adaptively casted into several groups according to their responses. This procedure will repeat until the convergent results are obtained.

\subsubsection{Adaptive Sampling and Classification}
The interested frequency band is first divided into several parts for evaluating the errors. Then, we select a small number of sampling points and find the extreme points according to their responses. They are then divided into several groups. In order to reduce the error in the whole frequency band, the maximum number of points in each group is limited to no more than 3. Fig. 1(b) presents an example with four groups including two non-single and single groups. 

\subsubsection{Interpolation} 
In our implementation, the Lagrange linear and quadratic interpolation are used. {$\bf{Algorithm~1}$} presents the procedure of calculating the approximate response at a desired frequency point. The group type should be checked before the interpolation. Since non-single groups do not include extreme points, the interpolation results are usually more accurate and stable. In order to ensure the convergence, the points with the distance not greater than ${{D}_{i}}/2$ at the endpoints are considered to be in the group ${{X}_{i}}$. 
\begin{algorithm}\small  
	\caption{Interpolation}  
	\begin{algorithmic}[1]
		\Require  {$X_i,x_{app},M_{ns},M_s$} 
		\Ensure   {$y_{app}$}        
		\For {$i=1 : M_{ns}$} 
		\If {$\left[\text{min}({X_i}) - \frac{D_i}{2} \right] \le x_{app} \le \left[\text{max}({X_i})+ \frac{D_{i+1}}{2}\right]$}
		\State isfind = true;
		\State $y_{app}$ = LagrangeInterpolation($X_i,x_{app}$);
		\EndIf 
		\EndFor
		\If{isfind = false} 
		\For{$i=1 : M_s$} 
		\If {$|x_{app} - X_i| < |\text{mindistance}|$}  
		\State mindistance = $x_{app} - X_i$;
		\State $X_{near} = X_i$;
		\EndIf 
		\EndFor 
		\State  $y_{app}$ = AdaptiveInterpolation($X_{near},x_{app}$); 
		\EndIf 
		\State {\bf{Return}} $y_{app}$
	\end{algorithmic}  
\end{algorithm}

For these points, which are not in the non-single groups, ${{X}_{near}}$ is used to construct the adaptive interpolation scheme, which are demonstrated in {$\bf{Algorithm~2-3}$}. {$\bf{Algorithm~2}$} presents the interpolation strategy for the scenario, where ${{X}_{near}}$ contains only an extreme point and ${{x}_{app}}$ is in its positive direction. For the points in its negative direction, a similar strategy can be used with only replacing ${{X}_{near+1}}$ by ${{X}_{near-1}}$ and changing ${{x}_{app-1}}>{{X}_{near}}$ with ${{x}_{app-1}}>{{X}_{near-1}}$.  $Case_{app}$ is used to record the chosen strategy for subsequent optimization. When ${{X}_{near}}$ does not include an extreme point, the procedure is introduced in {$\bf{Algorithm~3}$}. In this scenario, the former or latter group includes an extreme point.

\begin{algorithm}\small  
	\caption{Scenario \#1: Adaptive Interplation }  
	\begin{algorithmic}[1]
		\Require  {$X_{near},x_{app}$} 
		\Ensure   {$y_{app}$}        
		\If {$X_{near} \text{has an extreme point}, \text{mindistance} > 0$}
		\If{$X_{near+1} \neq \text{single group}$}
		\If{$x_{app - 1} > X_{near}$}
		\State $y_{app}$ = LagrangeInterpolation($x_{app-1},X_{near+1}^1,$
		\Statex \qquad \qquad  \qquad \qquad $X_{near+1}^2,x_{app}$);
		
		\Else
		\State $y_{app}$ = LagrangeInterpolation($X_{near},X_{near+1}^{1},$
		\Statex \qquad \qquad  \qquad \qquad$X_{near+1}^2,x_{app}$);
		\State $Case_{app}$ = 2;
		\EndIf 
		\Else
		\If{$x_{app - 1} > X_{near}$}
		\State $y_{app}$ = LinearInterpolation($x_{app-1},X_{near+1}^1,x_{app}$);
		\State ${Case_{app}}$ = 3;
		\Else
		\State $y_{app}$ = LinearInterpolation($X_{near},X_{near+1}^1,x_{app}$);
		\EndIf
		\EndIf
		\EndIf   
		\State {\bf{Return}} $y_{app}$
	\end{algorithmic}  
\end{algorithm}

\begin{algorithm}\small  
	\caption{Scenario \#2: Adaptive Interpolation}  
	\begin{algorithmic}[1]
		\Require  {$X_{near},x_{app}$} 
		\Ensure   {$y_{app}$}        
		\If {$X_{near}  \text{has no extreme points}$}
		\If{$X_{near - 1} \neq \text{single group}$}
		\If{$x_{app - 1} > X_{near - 1}$}
		\State $y_{app}$ = LagrangeInterpolation($X_{near-1}^N,x_{app-1},$
		\Statex \qquad \qquad  \qquad \qquad$X_{near},x_{app}$);
		\Else
		\State $y_{app}$ = LagrangeInterpolation($X_{near-1}^{N-1},X_{near-1}^{N},$
		\Statex \qquad \qquad  \qquad \qquad$X_{near},x_{app}$);
		\EndIf 
		\Else
		\If{$X_{near + 1} \neq \text{single group}$}
		\If{$x_{app - 1} > X_{near}$}
		\State $y_{app}$ = LagrangeInterpolation($x_{app-1},X_{near+1}^1,$
		\Statex \qquad \qquad  \qquad \qquad$X_{near+1}^2,x_{app}$);
		\Else
		\State $y_{app}$ = LagrangeInterpolation($X_{near},X_{near+1}^1,$
		\Statex \qquad \qquad  \qquad \qquad$X_{near+1}^2,x_{app}$);
		\EndIf 
		\Else
		\State $y_{app}$ = LinearInterpolation($X_{near}$,$X_{near\pm1}$,$x_{app}$);
		\EndIf
		\EndIf
		\EndIf
		\State {\bf{Return}} $y_{app}$
		
	\end{algorithmic}  
\end{algorithm}
Another essential step is to optimize the previously calculated results since larger errors occur near the extreme points. The responses in the adjacent non-single group are used to optimize the results near the extreme points to improve the accuracy. For instance, if ${{x}_{app}}$ is in a non-single group and $Cas{{e}_{app-1}}=2$, we can recalculate the Lagrange polynomials with $X_{near}^{N}, {{x}_{app}}, X_{near+1}^{1}$ to get the corrected value ${y_{app-1}}$ at ${x_{app-1}}$. As for the linear interpolation, take $Cas{{e}_{app}}=3$ for example, if ${x_{app-1}}>X_{near+1}^{1}$, ${y_{app}}$ is calculated through the linear extrapolation of ${x_{app-1}}$ and $X_{near+1}^{1}$. Otherwise, the linear extrapolation of ${{X}_{near}}$ and $X_{near+1}^{1}$.

\subsubsection{Error Estimate}
If we obtain the responses of  $n+1$ frequency points, to quantitatively measure the performance, the relative error is defined as ${\sum\limits_{i=0}^{n+1}{\left| y_{i}^{\operatorname{int}}-\tilde{y}_{i}^{\operatorname{int}} \right|}} / {\sum\limits_{i=0}^{n+1}{\left| {y}_{i}^{\operatorname{int}} \right|}}$, where $\tilde{y}_{i}^{int}$ is the previously approximated results, $y_{i}^{int}$ is the refined results. The error is used to control the iteration procedure in the proposed scheme. 

\subsubsection{Error Control}
After completing one interpolation process, the relative errors in each frequency range  are calculated. In order to increase the convergence and reduce the computational overhead, only the frequency bands, which have larger relative error than the predefined threshold, are refined until the convergent results are obtained. 

\section{ERROR ANALYSIS}
If we have the $n+1$ sampling points ${{x}_{j}}$ together with frequency responses ${{f}_{j}}, j=0,\ldots ,n$. The Lagrange’s interpolation formulation can be expressed as 
\begin{equation}\label{Function}
	g(x)=\sum\limits_{i=0}^{n}{{{f}_{i}}{{v}_{i}}(x)},
\end{equation}
where ${v}_{i}(x)$ is the Lagrange polynomial associated to ${x}_{i}$, which can be expressed as ${{v}_{i}}(x)={ \underset{k=0,k\ne i}{\overset{n}{\mathop{\prod }}}\, \frac{\left( x-x_k \right)}{\left( x_i-x_k\right)}}$.

Using the Taylor series analysis, the truncation error can be expressed by

\begin{equation}\label{Truncation}
	{{R}_{n}}(x)=f(x)-g(x)=\frac{{{f}^{(n+1)}}(\xi )}{(n+1)!}v(x),
\end{equation}
where ${{f}^{(n+1)}}(\xi)$ is the $(n+1)$th derivative of $f$ at $\xi$, and $v(x)=(x-{{x}_{0}})(x-{{x}_{1}})\cdots (x-{{x}_{n}})$. If ${{f}^{(n+1)}}(\xi ) = 0$, the truncation error reduces to zero. We denote $R_{n}^{1}(x)=v(x)/(n+1)!$. If the sampling points are even and the interval is $h$, $R_{n}^{1}(x)$ is bounded by${{\left[ \frac{\pi }{2}(n+1) \right]}^{-\frac{1}{2}}} {{\left( \frac{h}{2} \right)}^{n+1}}$ \cite{Bound}.

By using the differentiation theorem for the Fourier transforms \cite{Fourier} and  
applying the Schwartz inequality, we obtain
\begin{align}\label{f3}
{{\left| {{f}^{(n+1)}}(x) \right|}^{2}}&\le \underbrace{\int_{-{{f}_{0}}}^{{{f}_{0}}}{{{\left| {{\omega}^{n+1}}\exp \left[ j\left( \omega x+(n+1)\frac{\pi }{2} \right) \right] \right|}^{2}}}df}_{\text{Term1}}  \notag \\ 
& +\underbrace{\int_{-{{f}_{0}}}^{{{f}_{0}}}{|}F(f){{|}^{2}}df}_{\text{Term2}}. 
\end{align}
where $\omega = 2\pi f$, and ${{f}_{0}}$ denotes the cut-off frequency. 
%
The first term is bounded by 
\begin{equation}\label{f4}
	\begin{aligned}
	{{\left| {\text{Term1}}  \right|}}&\le {{(2\pi )}^{2n+2}}\int _{-{{f}_{0}}}^{{{f}_{0}}}{{f}^{2n+2}}df\\
	&={{\left[\pi (2n+3) \right ]}^{-1}}{{\left( 2\pi {{f}_{0}} \right)}^{2n+3}}. 
	\end{aligned}
\end{equation}
The second term is known as the finite energy of $f$, which can be represented by ${{B}^{2}}$. By substituting (\ref{f4}) into (\ref{f3}), we obtain
\begin{equation}\label{f5}
	{\left| {{f}^{(n+1)}}(x) \right|} \le B\pi {{(2n+3)}^{-\frac{1}{2}}}{{\left( 2\pi {{f}_{0}} \right)}^{n+\frac{3}{2}}}.
\end{equation}
 
By combining (\ref{f5}) and (\ref{Truncation}), we obtain
\begin{equation}\label{f6}
	{{R}_{n}}(x)\le \left| \frac{B}{n+1} {{\left( \frac{2}{\pi }{{f}_{0}} \right)}^{\frac{1}{2}}} {{\left( \pi h{{f}_{0}} \right)}^{n+1}} \right|.
\end{equation}

Therefore, if sufficient frequency points are selected, where the interval meets $h<{{\left( \pi {{f}_{0}} \right)}^{-1}}$, and the energy of $f$ is finite, the convergent results can be obtained. In addition, as the number of sampling points increases, the error becomes smaller.  

\section{NUMERICAL RESULTS AND DISCUSSION}
\subsection{Antenna Radiation}
The first numerical example is to calculate the responses of the Potter Horn Antenna, which is a dual-mode feedhorn and provides excellent radiation pattern with suppressed sidelobes and symmetrical beamwidths \cite{AntennaRes}. The frequency band is 27 GHz-33 GHz. If the frequency band is divided into 70 parts and the predefined relative error in each part is set to 0.03, 130 frequency points are selected to calculate and the responses in other frequency points can be evaluated through the proposed interpolation scheme. The average error in the whole frequency region is converged to 0.0019. Fig. 2 shows the results obtained from the HFSS, the proposed scheme in this case. It is easy to find that the results obtained from the proposed interpolation scheme show excellent agreement with those obtained from the HFSS both in the discrete and interpolation sweep modes. The frequency interval is 0.01 GHz and the number of points is 601 in the discrete mode. However only 130 points need to be calculated directly in the proposed scheme, which accounts for 21.6\% of the number of frequency points in the discrete mode. 

\begin{figure}[H]
	\begin{minipage}[]{0.96\linewidth}
		\centering
		\centerline{\includegraphics[width=2.50in]{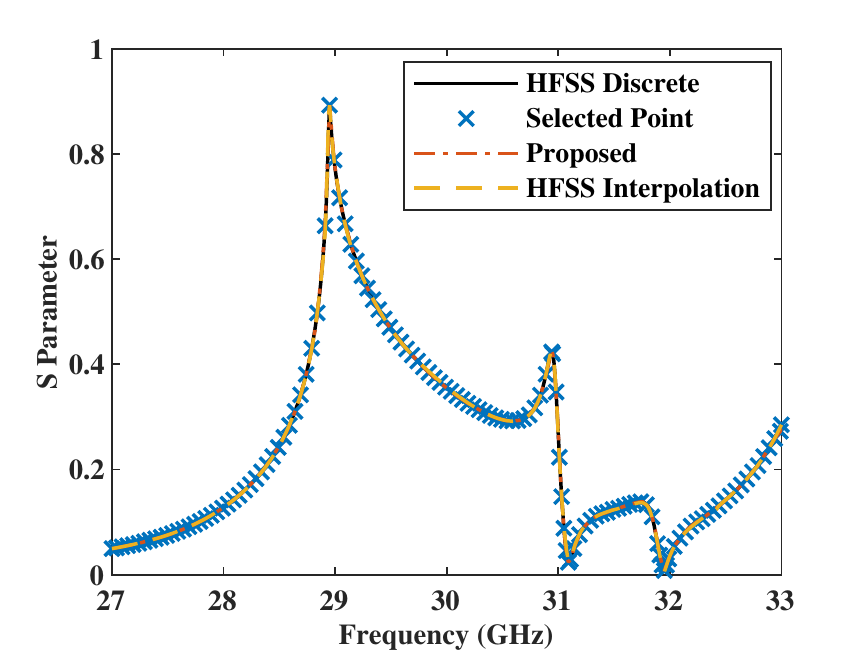}}		
		\label{FIG2}
	\end{minipage}
	\caption{The S Parameter obtained from the HFSS in the discrete mode, the HFSS in the interpolation mode, and the proposed interpolation scheme.}
\end{figure}

Fig. 3 (a) shows the interpolation error and the number of points versus the number of parts, the error in each part is set to 0.03. As the frequency band divided into different parts, different number of points are needed, with more points calculated, the smaller errors are usually obtained. In this numerical example, if the frequency range is divided into 70 parts, it can be found that the smallest error is achieved, with 130 points are used. Fig. 3 (b) shows the interpolation error and the number of points versus the error limit. It can be found that as the predefined error increases, the number of points required to be calculated decreases.

\begin{figure}[H]
	\begin{minipage}[h]{0.48\linewidth}
		\centering
		\centerline{\includegraphics[width=1.80in]{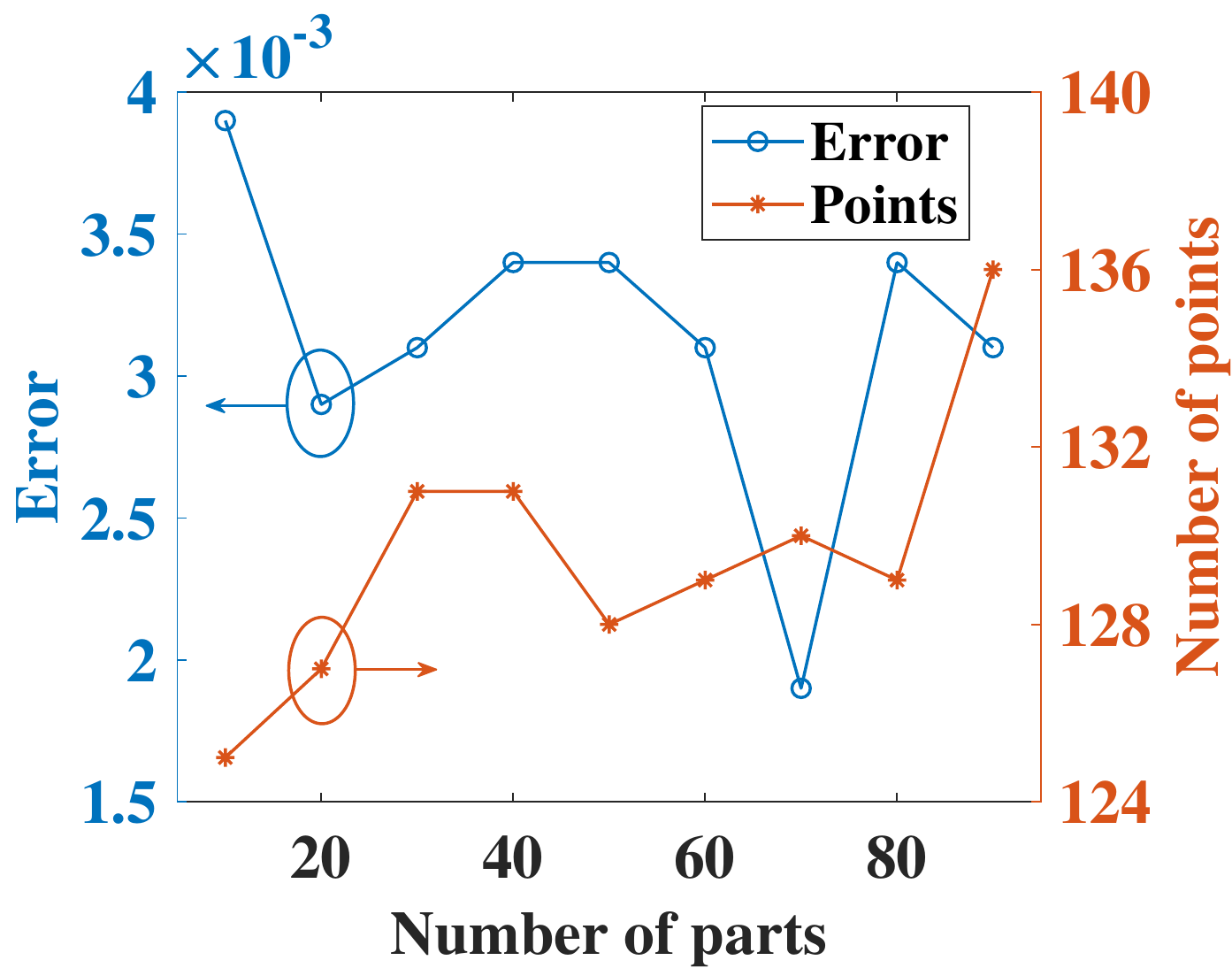}}		
		\centerline{(a)}
		\label{FIG1A}
	\end{minipage}
	\hfill
	\begin{minipage}[h]{0.48\linewidth}\label{FIG1B}
		\centerline{\includegraphics[width=1.80in]{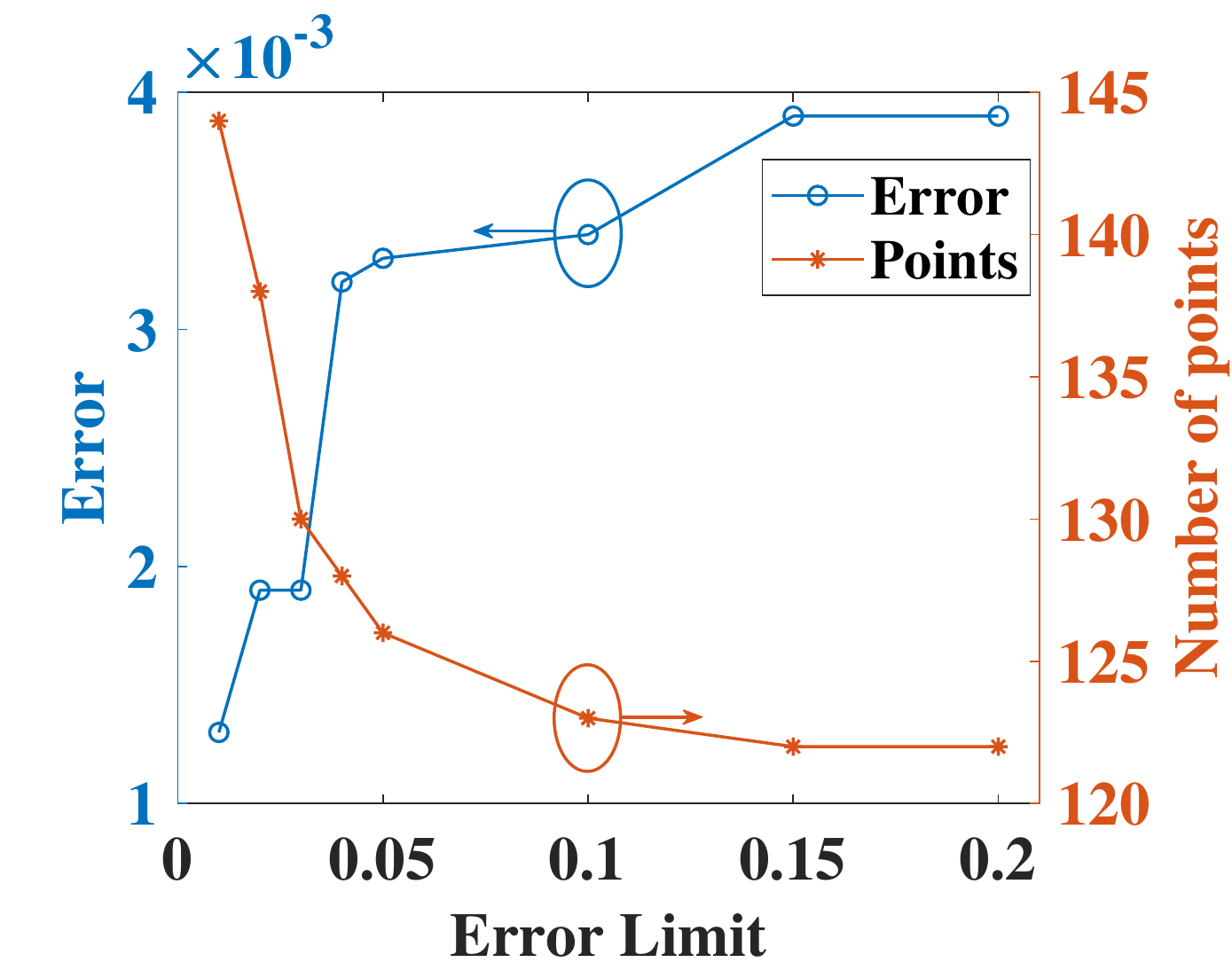}}
		\centerline{(b)}
	\end{minipage}
	\caption{(a) The error and the number of points with the error limit in each part set to 0.03, (b) the error and the number of points with the frequency band divided into 70 parts.}
	\label{fig_6}
\end{figure}

\subsection{The S Parameters of A Filter}
In order to further validate the accuracy of the proposed scheme with large variations, we consider an interdigital bandpass filter with a 1 GHz bandwidth. The S parameter from 0.6 GHz to 2.4 GHz is considered, and 601 points with an interval of 0.003 GHz is used in our simulations. When the frequency band is divided into 70 parts and the relative error in each part is limited to 0.05, only 129 discrete points are required to be calculated in the  proposed interpolation scheme. It accounts for only 21.5\% of the desired frequency points. The global error in the whole frequency band converges to 0.012. Fig. 4 shows the results obtained from the HFSS and the proposed interpolation scheme. Similar to the previous numerical example, as a result of the adaptive sampling strategy, the extreme points can be selected and the proposed scheme shows excellent agreement with those from the HFSS in the discrete and interpolation modes.

\begin{figure}[H]
	\begin{minipage}[]{0.96\linewidth}
		\centering
		\centerline{\includegraphics[width=2.50in]{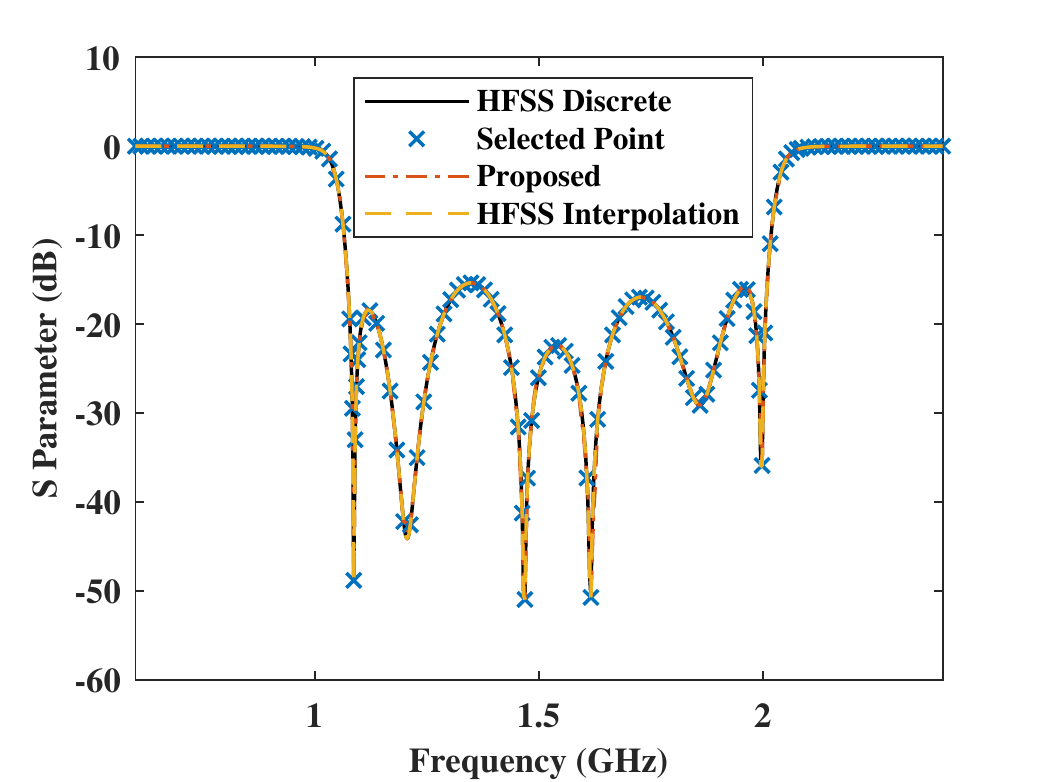}}		
		\label{FIG3}
	\end{minipage}
	\caption{The S Parameter obtained from the HFSS in the discrete mode, the HFSS in the interpolation mode, and the proposed interpolation scheme.}
\end{figure}

Fig. 5 (a) shows the interpolation error and the number of points with the frequency band divided into different parts when the relative error in each part is limited to 0.03. In Fig. 5 (b), the interpolation results are listed with the frequency band divided into 70 parts and the relative error in each part limited to be different values. As we can see, the observations are similar to the previous example. As the relative error increases, less points are required to be calculated. 

\begin{figure}[H]
	\begin{minipage}[h]{0.48\linewidth}
		\centering
		\centerline{\includegraphics[width=1.80in]{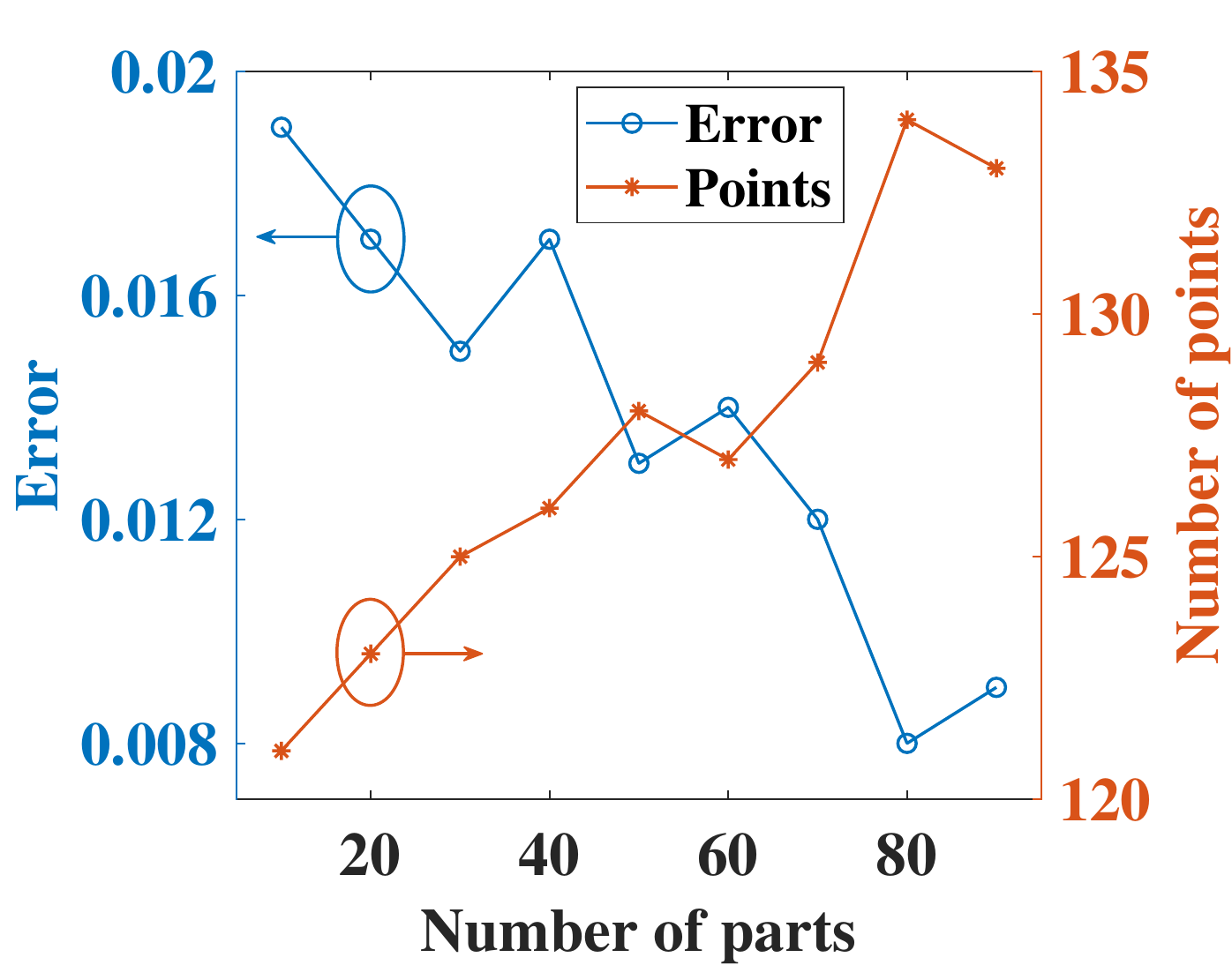}}		
		\centerline{(a)}
		\label{FIG1A}
	\end{minipage}
	\hfill
	\begin{minipage}[h]{0.48\linewidth}\label{FIG1B}
		\centerline{\includegraphics[width=1.80in]{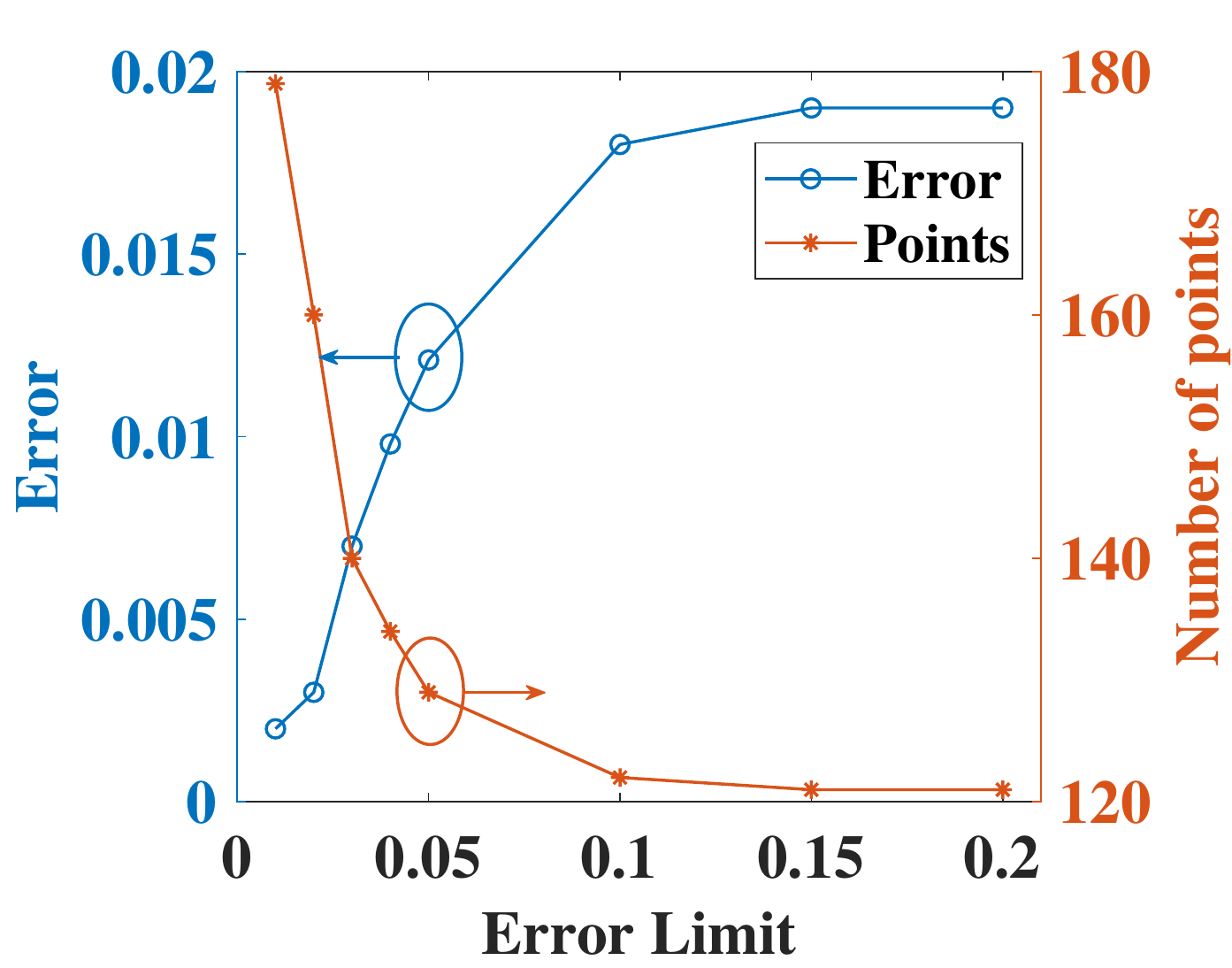}}
		\centerline{(b)}
	\end{minipage}
	\caption{(a) The error and the number of points with the error limit in each part set to 0.03, (b) the error and the number of points with the frequency band divided into 70 parts.}
	\label{fig_6}
\end{figure}

\section{Conclusion}
An adaptive interpolation scheme with a controllable error for fast frequency sweep in a wide frequency band is proposed. In the proposed scheme, we only need to calculate a small number of frequency points, and then adaptively use different interpolation strategies for other required frequency points. Numerical examples show that the proposed scheme can produce very accurate results and reveals its fast frequency sweep capability. Moreover, the proposed scheme can be used to calculate other complex responses in a wide variable range without being limited to the specific problems.

%




\end{document}